\documentclass[preprint,preprintnumbers,nofootinbib,aps,10pt,twocolumn]{revtex4-1}
\usepackage{amsmath,amssymb,bm,epsfig}
\usepackage{color}
\usepackage{natbib}
\usepackage{hyperref}
\usepackage{ulem}
\usepackage{graphicx}
\usepackage{xcolor,colortbl}
\RequirePackage{lineno}
\newcommand{\nn}{\nonumber}
\newcommand{\sNN}{\sqrt{s_{\textrm{NN}}}}

\definecolor{Gray}{gray}{0.85}
\newcolumntype{a}{>{\columncolor{Gray}}c}

\def \beq{\begin{equation}}
\def \eeq{\end{equation}}
\def \beqa{\begin{eqnarray}}
\def \eeqa{\end{eqnarray}}

\begin{document}

\title{Effect of hadronic interaction on the flow of $K^{*0}$}

\author{Tribhuban Parida}
\email{tribhubanp18@iiserbpr.ac.in}
\author{Sandeep Chatterjee}
\email{sandeep@iiserbpr.ac.in}
\author{Md. Nasim}
\email{nasim@iiserbpr.ac.in}

\affiliation{Department of Physical Sciences,\\
Indian Institute of Science Education and Research Berhampur,\\ 
Transit Campus (Govt ITI), Berhampur-760010, Odisha, India}

\begin{abstract}
We explore the implications of the late stage hadronic 
rescattering phase on 
the flow of $K^{*0}$. The model calculations are done 
using a (3+1)-dimensional hybrid framework, incorporating
both hydrodynamic evolution and hadronic transport that 
is calibrated to agree with bulk observables including 
the elusive rapidity differential $v_1$  of light-flavor
hadrons. We find that the late stage hadronic rescattering phase causes significant qualitative modification of the $K^{*0}$ $v_1$ resulting in $\frac{dv_1}{dy}(K^{*0})-\frac{dv_1}{dy}(K^{+})$ and $\frac{dv_1}{dy}(\phi)-\frac{dv_1}{dy}(K^{+})$ to have opposite signs with the effect being more pronounced in central 
collisions as compared to peripheral ones due to the larger multiplicity as well as 
longer duration of the hadronic phase. Further, this effect is enhanced in low-energy 
collisions owing to a stronger breaking of boost 
invariance. On the contrary, the influence of the hadronic phase on the $K^{*0}$ elliptic 
flow $v_2$ is found to be less significant and quantitative.

\end{abstract}

\maketitle

\section{Introduction}

The effect of the late stage hadronic rescattering phase 
on the yield of resonances have been well studied.
Short lived resonances like $\rho^{0}(770)$, $K^{*0}(892)$, $\Lambda^{*0}(1520)$ etc decay within the hadronic medium formed in the late stage of relativistic heavy ion collsions~\cite{ALICE:2018ewo,ALICE:2019smg,ALICE:2018qdv,STAR:2004bgh,ALICE:2014jbq,Song:2017hlr}. The resultant daughter particles undergo rescattering with other hadrons in the medium which inhibits the reconstruction of the resonace signal 
in experimental analysis~\cite{ALICE:2018ewo,ALICE:2019smg,ALICE:2018qdv,STAR:2004bgh,ALICE:2014jbq,Song:2017hlr, ALICE:2019xyr,STAR:2010avo,Bleicher:2002dm}. This results in a notable reduction in the final yield of these resonances. On the other hand, pseudo-inelastic interaction between the hadrons in the medium can increase the resonance yield through regeneration process~\cite{Bleicher:2002dm},  These hadronic state effects on resonances is experimentally verified by measuring the non-resonance to resonace yield ratio across systems of varying sizes~\cite{STAR:2004bgh,STAR:2006vhb,ALICE:2018qdv,ALICE:2014jbq,ALICE:2018ewo,ALICE:2017pgw}. The change in the yield of resonances depends on both the hadronic phase lifetime and density of hadrons in the medium~\cite{Oliinychenko:2021enj,Sahoo:2023rko,LeRoux:2021adw,Werner:2018yad,Knospe:2021jgt}. This distinctive behavior positions the resonances as ideal candidates for probing the hadronic phase of heavy-ion collisions. 

The $K^{*0}$ resonance has a short lifetime of $\sim 4$ fm/$c$. It deacys to $\pi$ and $K$ in the medium. The resultant daughter pions mainly undergo scattering with the hadronic medium of the fireball 
that is dominantly pions leading to loss of signal of $K^{*0}$. However, the regeneration of $K^{*0}$ is less pronounced due to the fact that the $\pi-K$ cross-section is smaller than that of $\pi-\pi$ interactions~\cite{Protopopescu:1973sh,Matison:1974sm,Bleicher:1999xi,STAR:2004bgh}. Consequently, this gives rise to a reduction in the $K^{*0}$ to $K$ yield ratio in larger systems compared to p+p collisions at equivalent collision energies. This distinguishing characteristic of the $K^{*0}$ is used to get a rough estimate of the time span between chemical and kinetic freeze-outs in a given system~\cite{Motornenko:2019jha,STAR:2022sir}. Such in-medium effects can also influence the phase space distribution of the $K^{*0}$ resonance, which can in turn be reflected in the flow coefficients. 
The flow coefficients are characterized by different order harmonics in the Fourier series expansion of the azimuthal distribution of particles produced in momentum space
\beq
\frac{dN}{p_T dp_T dy d\phi} = \frac{dN}{p_T d p_T dy} \left( 1 + 2 \sum_{n=0}^{\infty} v_{n}(p_T, y) \cos{ \left[ n(\phi-\psi_{n}) \right] } \right)
\eeq
Here, $\psi_n$ is the event plane angle associated with the $n^{th}$ order harmonics. The variables $p_T$, $y$, and $\phi$ represent the transverse momentum, rapidity, and azimuthal angle of the produced particles, respectively.

It is known from earlier studies that the hadronic interactions affect the elliptic flow ($v_2$) of $K^{*0}$ at low $p_T$~\cite{Oliinychenko:2021enj}. In the current study, we focus on the rapidity-odd directed flow ($v_1$) of $K^{*0}$. The study of the $v_1$ of various light flavor hadrons has been conducted to provide deeper understanding of several key aspects. These include constraining the initial three-dimensional distribution of energy and baryon density in the medium~\cite{Bozek:2010bi,Ryu:2021lnx,Jiang:2021ajc,Shen:2020jwv,Parida:2022ppj,Parida:2022zse,Bozek:2022svy,Jiang:2023fad}, investigating the characteristics of the QCD equation of state~\cite{Steinheimer:2014pfa,Rischke:1995pe,Nara:2016phs,Ivanov:2016spr,Ivanov:2014ioa,Steinheimer:2014pfa}, and extracting the transport coefficients of the medium~\cite{Parida:2023rux,Parida:2022ppj,Mohs:2020awg,Becattini:2015ska}.
The study of $v_1$ of the $K^{*0}$ resonance can provide information about the extent to which these resonances participate in the collective expansion of the system. Also the impact of the hadronic afterburner on $v_1$ can yield insights into the coordinate as well as momentum space configuration of the late stage hadronic fireball.

In the next section we describe the framework that has been used in this study. The results are presented and explained in Sec.\ref{result_sec} and we will summarize the findings in Sec.\ref{summ_sec}.

\section{Framework}
The framework used in this study includes multiple components to simulate different stages of the heavy ion collisions. 
A Glauber based model has been used to set up the initial condition for the hydrodyanmic evolution. 
The expansion of the resulting fireball is simulated by the publicly available MUSIC 
code~\cite{Schenke:2010nt,Schenke:2011bn,Paquet:2015lta,Denicol:2018wdp}. 
The iSS code~\cite{Shen:2014vra,https://github.com/chunshen1987/iSS} has been used to sample the primordial hadrons from the hypersurface of constant energy density, generated 
from the space-time evolution of the fluid. Subsequently, the UrQMD code~\cite{Bass:1998ca,Bleicher:1999xi} is employed to simulate the interaction and expansion 
of the hadrons during the dilute phase of the heavy ion collision.

Smooth transverse profiles of participant and binary collision densities has been prepared by averaging over 25,000 MC Glauber events~\cite{Shen:2020jwv}. We have set the impact parameter direction along x-axis in each event. 
The participant and binary collision sources obtained from each MC Glauber event are rotated by the second-order participant plane angle 
and then smeared out in the transverse plane. The smearing profile is assumed to be a Gaussian with parametric width $\sigma_{\perp} = 0.4$ fm.    
Using the transverse profiles of participant and binary collision densities, we have constructed the initial profile of energy density at a constant proper time($\tau_{0}$) which takes the
following form.

\beqa
  \epsilon(x,y,\eta_{s}; \tau_{0}) &=& \epsilon_{0} \left[ \left( N_{+}(x,y) f_{+}(\eta_{s}) + N_{-}(x,y) f_{-}(\eta_{s})  \right)\right.\nn\\
                           &&\left.\times \left( 1- \alpha \right) + N_{bin} (x,y)  \epsilon_{\eta_s}\left(\eta_{s}\right) \alpha \right] 
 \label{eq.tilt}
\eeqa
where, $N_{+}(x,y)$  and $N_{-}(x,y)$ are the participant densities of the nuclei 
moving with positive and negative rapidity respectively. $N_{bin} (x,y)$ accounts for the contributions from binary collision 
sources at each point in the transverse plane. $\alpha$ 
is the hardness factor which controls the relative contribution of participant and binary sources in the total deposited
energy. The space-time rapidity($\eta_s$) extension profile, $\epsilon_{\eta_s}(\eta_s)$ is an even function of $\eta_s$ which has the following form.
\begin{equation}
  \epsilon_{\eta_s}(\eta_s) = \exp \left(  -\frac{ \left( \vert \eta_{s} \vert - \eta_{0} \right)^2}{2 \sigma_{\eta}^2}   
    \theta (\vert \eta_{s} \vert - \eta_{0} ) \right)
    \label{eq_etas_even_profile_for_epsilon}
\end{equation}
Here, $\eta_0$ and $\sigma_{\eta}$ are two free parameters which are tuned to capture the rapidity differential charged particle yield.

The functions $ f_{+,-}(\eta_s)$ introduce the assymetric deposition of matter in forward and
backward rapidity region.
\begin{equation}
    f_{+,-}(\eta_s) = \epsilon_{\eta_s}(\eta_s) \epsilon_{F,B}(\eta_s)
\end{equation}
with
\begin{equation}
    \epsilon_{F}(\eta_s) = 
    \begin{cases}
    0, & \text{if } \eta_{s} < -\eta_{m}\\
    \frac{\eta_{s} + \eta_{m }}{2 \eta_{m}},  & \text{if }  -\eta_{m} \le \eta_{s} \le \eta_{m} \\
    1,& \text{if }  \eta_{m} < \eta_{s}
\end{cases}
\label{tilt_prof}
\end{equation}
and 
\begin{equation}
    \epsilon_{B} (\eta_s) = \epsilon_F(-\eta_s)
\end{equation}
This deposition scheme creates a tilted profile of energy density in the reaction plane (the plane made by impact paramter and beam direction of the collision)~\cite{Bozek:2010bi} 
where the tilt is controlled by the model parameter $\eta_m$.

The baryon deposition scheme used in this work was first introced in~\cite{Parida:2022ppj}, where the initial baryon distribution depends 
on both participant and binary collision sources. The three dimensional distribution of the baryon density at $\tau_{0}$ is,
\begin{equation}
    n_{B} \left( x, y, \eta_s ; \tau_{0} \right) = 
       N_{B} \left[ W_{+}^{B}(x,y) f_{+}^{B}(\eta_{s}) + W_{-}^{B}(x,y) f_{-}^{B}(\eta_{s})  \right].
    \label{my_baryon_ansatz}
\end{equation}
$f_{\pm}^{n_{B}}$ are the rapidity envelope profiles for the net baryon deposition which are taken as~\cite{Denicol:2018wdp,Shen:2020jwv}, 
\beqa
    f_{+}^{n_{B}} \left( \eta_s \right) &=&  \left[  \theta\left( \eta_s - \eta_{0}^{n_{B} } \right)   \exp{- \frac{\left( \eta_s - \eta_{0}^{n_{B} }  \right)^2}{2 \sigma_{B, + }^2}}   + \right.\nn\\ && \left. \theta\left(  \eta_{0}^{n_{B} } - \eta_s \right)   \exp{- \frac{\left( \eta_s - \eta_{0}^{n_{B} }  \right)^2}{2 \sigma_{B, - }^2}}   \right]
\label{forward_baryon_envelop}
\eeqa
and
\beqa
    f_{-}^{n_{B}} \left( \eta_s \right) &=&  \left[   \theta\left( \eta_s + \eta_{0}^{n_{B} } \right)   \exp{- \frac{\left( \eta_s + \eta_{0}^{n_{B} }  \right)^2}{2 \sigma_{B, - }^2}}   + \right.\nn\\ && \left. \theta\left( -\eta_s -  \eta_{0}^{n_{B} }  \right)   \exp{- \frac{\left( \eta_s + \eta_{0}^{n_{B} }  \right)^2}{2 \sigma_{B, + }^2}}   \right]
\label{backward_baryon_envelop}
\eeqa
Here, $\eta_{0}^{n_{B}}$ and $\sigma_{B,\pm}$  are the model parameters which are tuned to capture the rapidity differential net proton yield.
$W_{\pm}^{B}(x,y)$ are the weight factors to deposit the net baryon in the transverse plane which has the following form.
\begin{equation}
W_{\pm}^{B}(x,y) = \left( 1 - \omega \right) N_{\pm}(x,y) + \omega N_{bin}(x,y)
    \label{weight_ansatz_1_for_baryon}
\end{equation}
The $N_B$ in Eq.~\ref{my_baryon_ansatz} is fixed by the constraint,
\beq
\int \tau_0 dx dy d\eta_s n_{B}\left(x, y, \eta_s ; \tau_{0} \right) = N_{\text{part}},
\eeq
where $N_{\text{part}} = \int dx dy \left[ N_{+}(x,y) + N_{-}(x,y) \right] $. The model parameter $\omega$ in Eq.~\ref{weight_ansatz_1_for_baryon} determines the proportionate contributions from participant and binary sources in 
the baryon profile. It has been observed that $\omega$ acts as a
tilt parameter for the baryon profile~\cite{Parida:2022ppj}. The value of $\eta_m$ and $\omega$ decides 
the relative tilt between energy and baryon density profile and a suitable choice of the value of these two parameter can describe 
the experimentally measured $v_1(y)$ of $\pi^{\pm},p$, $\bar{p}$ and other hadrons simultaneously~\cite{Parida:2022zse}. 

We start the hydrodynamic evolution of the energy and baryon density with zero initial transverse velocity by following the Bjorken flow ansatz.
A constant specific shear viscosity($C_{\eta} = \frac{\eta T}{\epsilon+p}=0.08$) has been taken during the fluid evolution at all collision energies. However, we have not considered the effect of bulk viscosity setting its value to zero. To introduce non-zero baryon diffusion, we utilize the 
following expression for the baryon transport coefficient ($\kappa_{B}$) which is derived from the Boltzmann equation in the relaxation time approximation~\cite{Denicol:2018wdp}.
\beq
\kappa_{B} = \frac{C_B}{T} n_{B} \left[ \frac{1}{3} \coth{\left(\frac{\mu_B}{T}\right)} - \frac{n_B T}{\epsilon + p} \right]
\eeq
In this equation, $C_B$ is a model parameter governing the strength of baryon diffusion in the medium which we have taken as unity. In this expression, $n_B$ represents the net baryon density, $p$ is the local pressure, $T$ corresponds to the temperature, and $\mu_B$ stands for the baryon chemical potential of the fluid. The Equation of State (EoS) utilized in this study enforces both strangeness neutrality and a fixed baryon-to-charge density ratio within each fluid cell~\cite{Monnai:2019hkn}.
The particlization has been performed on the hypersurface characterized by constant energy density($\epsilon_f = 0.26$ GeV/fm$^3$) 
using the iSS code~\cite{Shen:2014vra,https://github.com/chunshen1987/iSS}. From the hypersurface, we have generated a large number of particlization events and in each event the resultant hadrons are subsequently put into UrQMD~\cite{Bass:1998ca,Bleicher:1999xi} for late-stage hadronic interactions.

The value of the model parameters taken in this study at different $\sqrt{s_{NN}}$ are same as taken in Ref.~\cite{Parida:2022zse}. Notably, the model 
captures the experimental data of pseudorapidty differential charge particle yield, $p_T$ spectra of identified hadrons,  $p_T$ dependent $v_2$ of charged hadrons, rapidity differential
net proton yield and rapidity differential directed flow of identified hadrons with the chosen model parameters  ~\cite{Parida:2022ppj,Parida:2022zse}.

\section{Results}
\label{result_sec}
First, we have studied the directed flow of $K^{*0}$ and $\phi$ in Au+Au collisions of 10-40$\%$ centrality at $\sNN = 27$ GeV. In order to investigate the effect of late stage hadronic interaction, we
have considered two different scenarios of final particle production. In the first scenario, primordial resonances produced from the hypersurface 
instantaneously decay into stable hadrons without undergoing any further interactions. In contrast, the second scenario involves feeding the 
primordial hadrons into UrQMD. Subsequently, the system of hadrons evolves through a sequence of binary collisions following the Boltzmann 
equation. In this latter case, the hadrons undergo multiple instances of both elastic and inelastic scattering in the hadronic phase to give 
the ultimate stable hadrons. Furthermore, the directed flow of primordial hadrons is also presented to demonstrate the exclusive effect of 
resonance decay on directed flow.

\begin{figure}
 \begin{center}
  \includegraphics[scale=0.7]{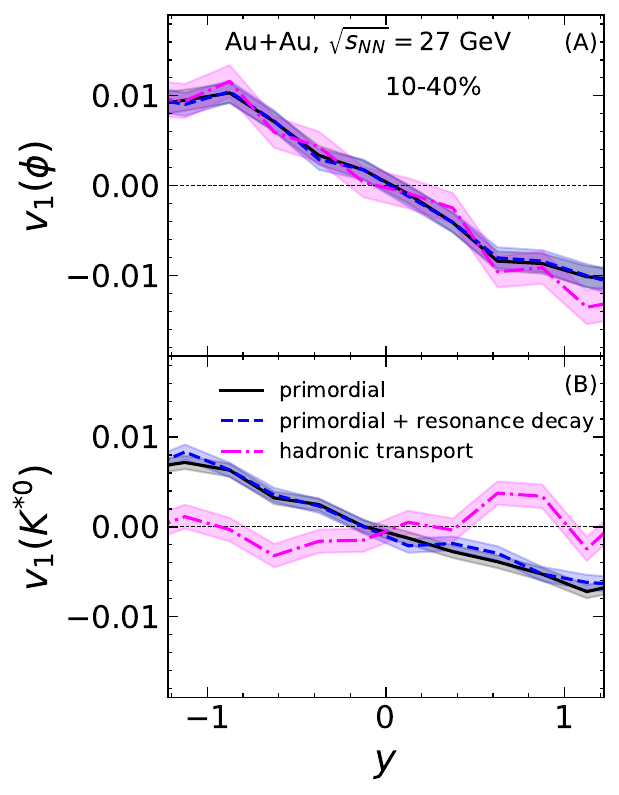}
 \caption{The rapidity differential directed flow($v_1$) of  $\phi$ and $K^{*0}$ has been plotted in panel (A) and (B) respectively for
 Au+Au collisons of 10-40$\%$ centrality at $\sqrt{s_{NN}} = 27$ GeV .
 The $v_1$ of the primordial hadrons which are produced directly from the hypersurface has been represented in solid lines. The dashed lines represent the $v_1$ calculations of the hadrons after performing the decay of resonances to stable hadrons whereas the dashed-dotted lines represents the $v_1$ calculations of the hadrons after they pass through the UrQMD hadronic afterburner. The band in each line provides the statistical uncertainty in the calculation. }
 \label{fig1}
 \end{center}
\end{figure}

The rapidity-dependent directed flow of $\phi$ and $K^{*0}$ are plotted in Figs.~\ref{fig1}(A) and 1(B) respectively. 
Due to the small scattering cross sections, the phase space distribution of the $\phi$ meson is less affected within the hadronic medium~\cite{Hirano:2007ei,Takeuchi:2015ana,Shor:1984ui}. Further, owing to a larger lifetime than the fireball lifetime, $\phi$ is not affected by rescattering effects. Consequently, the $v_1$ of $\phi$ remains largely unaltered even after traversing through hadronic transport. However, the impact of hadronic transport is notably pronounced in $K^{*0}$.
Remarkably, it is observed that the mid-rapidity slope of $v_1(y)$ for $K^{*0}$ changes sign from negative (prior to the hadronic interaction) to positive (after the hadronic interaction). 

\begin{figure}
 \begin{center}
  \includegraphics[scale=0.7]{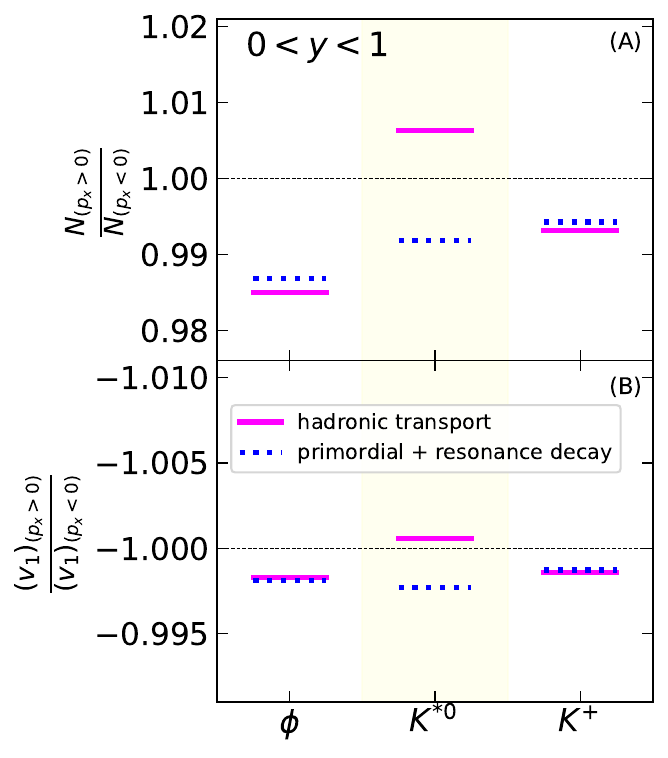}
 \caption{ The rapidity integrated yield (A) and $v_1$ (B) ratios between hadrons of $p_x>0$ and $p_x<0$ (see Eqs.~\ref{Eq.int_yld} and~\ref{Eq.int_v1}) in Au+Au collisons 
   of 10-40$\%$ centrality at $\sqrt{s_{NN}} = 27$ GeV. The rapidity integration has been performed between $0<y<1$. The ratios have been shown for $\phi$, $K^{*0}$ and $K^{+}$. To show the results obtained under different conditions, specifically, with or without the inclusion of hadronic transport stages for the hadrons, we have plotted the ratios for scenarios involving solely primordial hadrons and their resonance decay contributions (dotted lines), as well as for cases where the hadrons undergo interactions in the hadronic medium (solid lines).    }
 \label{fig2}
 \end{center}
\end{figure}

\begin{figure}
 \begin{center}
  \includegraphics[scale=0.7]{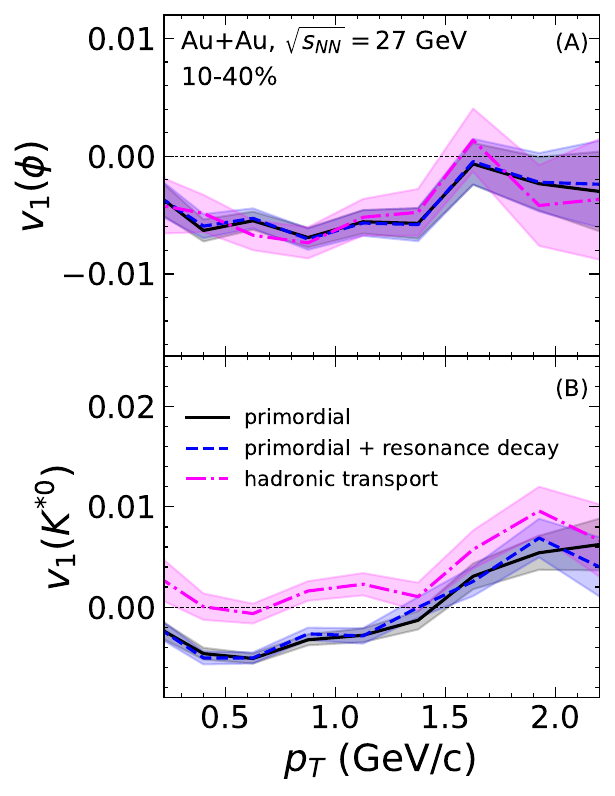}
 \caption{The $p_T$ differential directed flow ($v_1$) has been plotted for $\phi$ (A) and $K^{*0}$ (B) for
 Au+Au collisons of 10-40$\%$ centrality at $\sqrt{s_{NN}} = 27$ GeV. The calculation has been done for the particles produced within the rapidity range $0<y<1$. The $v_1$ of the primordial hadrons which are produced directly from the hypersurface has been represented in solid lines. The dashed lines represent the $v_1$ calculations of the hadrons after performing the decay of resonances to stable hadrons whereas the dashed-dotted lines represents the $v_1$ calculations of the hadrons after they pass through the UrQMD hadronic afterburner.}
 \label{fig3}
 \end{center}
\end{figure}

\begin{figure}
 \begin{center}
  \includegraphics[scale=0.7]{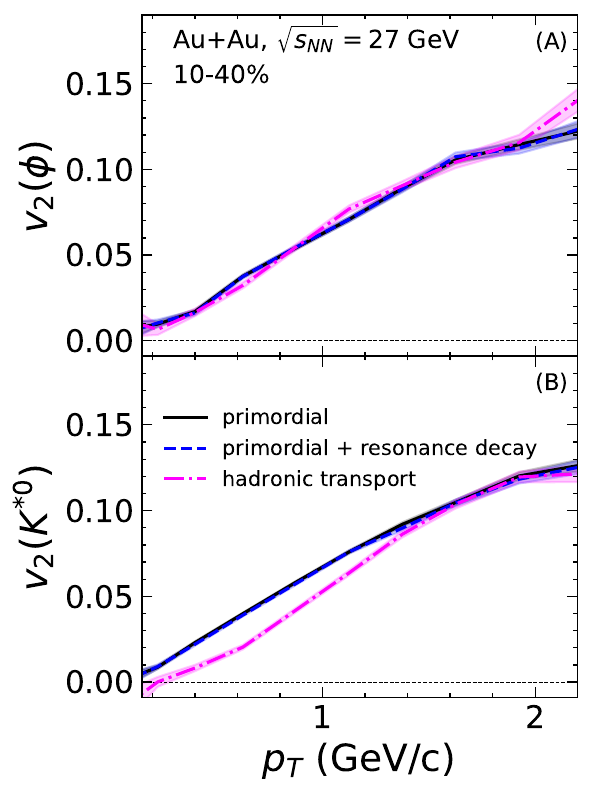}
 \caption{
 Same as in Fig.~\ref{fig3} but for
 the $p_T$ differential elliptic flow ($v_2$) of the hadrons produced within the rapidity range $\vert y \vert < 0.5$.}
 \label{fig.v2pt}
 \end{center}
\end{figure}

To understand the underlying reason behind the significant alteration in the $v_1$ of $K^{*0}$ during the hadronic phase, we have studied the variations in yield and $v_1$ for $K^{*0}$ within distinct regions of its phase space. Our focus remained exclusive to the positive rapidity region ($0 < y < 1$), where we independently examined the behavior of hadrons flowing with $p_x > 0$ and $p_x < 0$. The ratio of integrated yield ($ N$) and $v_1$ between hadrons with $p_x > 0$ and $p_x < 0$ is calculated
as follows: 
\begin{equation}
 \frac{N_{(p_x > 0)} }{N_{(p_x < 0)}} = \frac{ \int_{0}^{1} dy 
 \int_{-\pi/2}^{\pi/2} d\phi \int dp_T  \frac{dN}{ dp_T dy d\phi} } { \int_{0}^{1} dy 
 \int_{\pi/2}^{3 \pi/2} d\phi \int dp_T  \frac{dN}{ dp_T dy d\phi} }
 \label{Eq.int_yld}
\end{equation}
\begin{equation}
 \frac{(v_1)_{(p_x > 0)} }{(v_1)_{(p_x < 0)}} =  \frac{ \int_{0}^{1} dy 
 \int_{-\pi/2}^{\pi/2} d\phi \cos{\phi} \int dp_T  \frac{dN}{ dp_T dy d\phi} } { \int_{0}^{1} dy 
 \int_{\pi/2}^{3 \pi/2} d\phi \cos{\phi} \int dp_T  \frac{dN}{ dp_T dy d\phi} } \times  \frac{N_{(p_x < 0)} }{N_{(p_x > 0)}}
 \label{Eq.int_v1}
\end{equation}
The comparison of the rapidity integrated yield and $v_1$ ratios between hadrons with positive and negative $p_x$ for $\phi$, $K^{*0}$ and $K^{+}$ is depicted in Figs.~\ref{fig2}(A) and~\ref{fig2}(B) respectively. Within the plot, we present a comparison of outcomes obtained with and without the inclusion of hadronic transport stages for the hadrons.

It has been observed that the ratio of integrated yield for $K^{+}$ between $p_x > 0$ and $p_x < 0$ is less than one, and this value changes a little during the hadronic interactions. Similarly, the ratio of the magnitude of average directed flow ($ v_1 $) also shows a minute change in the hadronic phase and continues to be less than one. This implies that more $K^{+}$ particles are produced with negative $p_x$, leading to the observed negative value of $v_1$ in the positive rapidity region, which remains unaltered throughout the late-stage evolution.
The results are similar for $\phi$ as it mostly decays outside the fireball owing to a larger lifetime and hence there is minimal rescattering effect from the afterburner.
However, a significant effect of hadronic afterburner has been noted in the ratio of integrated $K^{*0}$ yield between particles with $p_x > 0$ and $p_x < 0$. After the hadronic interactions, the ratio, $N_{(p_x>0)}/N_{(p_x<0)}$ becomes greater than one, which contrasts the scenario where it was initially less than one in the absence of hadronic interactions. This change in the integrated yield ratio during the hadronic phase can be attributed to the distinct hadronic interactions experienced by $K^{*0}$ particles in different regions of momentum space due to initial tilted condition resulting in an asymmetric distribution of the hadronic fireball both in coordinate and momentum space.

Due to its short lifetime, the $K^{*0}$ resonance decays into daughter particles ($\pi$ and $K$) within the hadronic medium. If any of the daughter particles undergoes a change in momentum due to interactions with other particles in the medium, it becomes unfeasible to reconstruct the $K^{*0}$ using this rescattered daughter particle. This phenomenon is referred to as the "signal loss" of the $K^{*0}$~\cite{STAR:2022sir,STAR:2004bgh,STAR:2010avo,Li:2022neh,ALICE:2012pjb,ALICE:2021xyh}. In Fig.~\ref{fig2}(A), we observe a higher signal loss for $K^{*0}$ with $p_x<0$ which gives rise to $N_{(p_x>0)}/N_{(p_x<0)}>1$. The asymmetric signal loss on different sides of $p_x=0$ stems from the unequal distribution of pions on the two sides of the beam axis in the reaction plane. The initial tilted profile leads to an asymmetric evolution of the fireball, whereby a predominantly large number of pions flow with negative $p_x$~\cite{Bozek:2010bi,Jing:2023zrh,Jiang:2021ajc} resulting in an asymmetric distribution of the bath in coordinate space. As a result, $K^{*0}$ particles with $p_x<0$ located in the positive rapidity regions encounter a relatively denser pion medium. Consequently, a comparatively higher amount of rescattering occur, contributing to a greater loss of $K^{*0}$ signal in this region. Since we are capable of reconstructing more hadrons with $p_x>0$, we obtain a $v_1$ that is more weighted by $K^{*0}$ with $p_x>0$. This leads to an overall positive $v_1$ for $K^{*0}$ in the positive rapidity region. Due to the symmetry of the collision, 
the same mechanism makes the $v_1$ value of $K^{*0}$ negative in the negative rapidity region.

The $p_T$-dependent $v_1$ values for both $\phi$ and $K^{*0}$ are presented in Fig.~\ref{fig3}(A) and~\ref{fig3}(B) respectively. This $p_T$ differential calculation has been done for particles produced in positive rapidity ($y$) within range $0<y<1$. 
The afterburner has negligible effect on $v_1$ of $\phi$ as expected.
However, in the case of $K^{*0}$, a pronounced influence of hadronic interactions becomes evident, particularly at low $p_T$. The $v_1$ values of $K^{*0}$ resonance at relatively larger $p_T$ region ($p_T > 1.5$ GeV/c) is already positive at the hadronisation surface which indicates they are on the less dense side of the fireball and hence their decay products are likely to rescatter less       resulting in reduced effect on the $v_1$. Further, they are more likely to decay outside the fireball and hence the daughters escape rescattering by the hadronic medium allowing for reconstruction of $K^{*0}$. 

The effect of hadronic interaction on the resonances is also reflected in the elliptic flow~\cite{Oliinychenko:2021enj}. In Fig.~\ref{fig.v2pt}, we have plotted the $v_2$ as a function of $p_T$ for $\phi$ and $K^{*0}$ resonances generated within the mid-rapidity, $\mid y \mid < 0.5 $ region. Notably, it has been observed that the elliptic flow magnitude of $K^{*0}$ decreases for $p_T<1.5$ GeV/c as there is likely to be larger $K^{*0}$ loss
along in-plane direction than out of the reaction plane direction~\cite{Li:2022neh}. However, this suppression reduces as one goes to higher $p_T$  consistent with the observations in the $v_1$ case. Conversely, the impact of hadronic transport on the $v_2$ of $\phi$ is minimal.

\begin{figure}
 \begin{center}
  \includegraphics[scale=0.5]{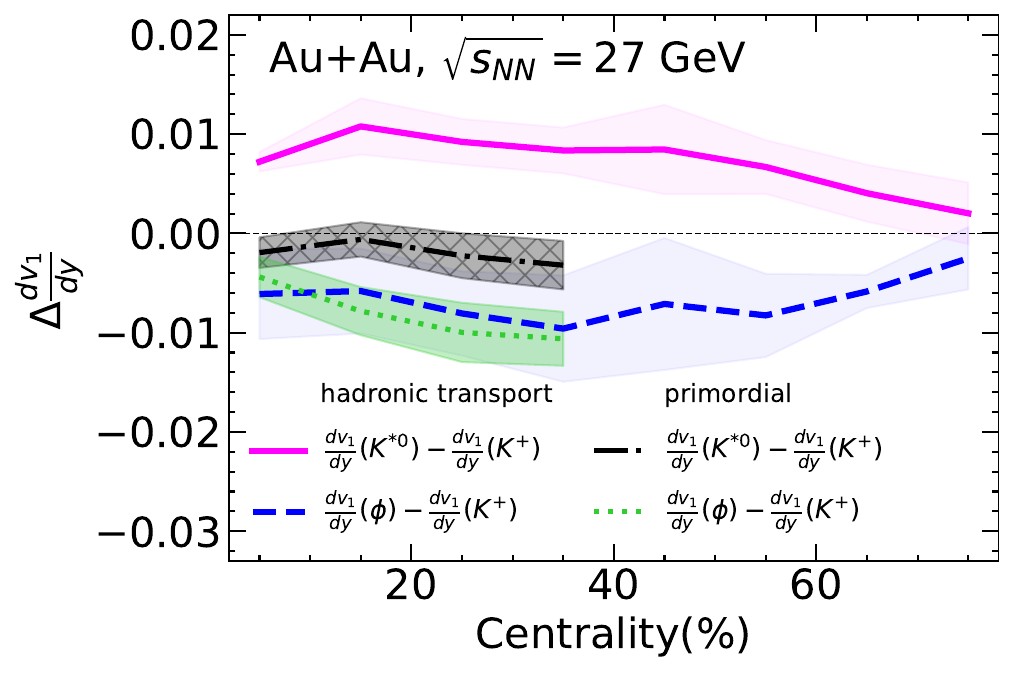}
 \caption{The splitting of directed flow slopes ($\frac{dv_1}{dy}$) of resonances $\phi$ and $K^{*0}$ with $K^{+}$ has been plotted as a function of centrality for Au+Au collisions at $\sqrt{s_{NN}} = 27$ GeV. The shaded bands denote the statistical uncertainties in the model calculations. }
 \label{fig4}
 \end{center}
\end{figure}

Our observation highlights a significant difference in the impact of hadronic interactions on the $v_1$ of $\phi$ and $K^{*0}$ particles. This difference becomes particularly interesting when we study their dependence on centralities and collision energies. Such an investigation could provide deeper insights into the role played by the hadronic phase in heavy ion collisions. In this regard, we present the centrality dpendence of mid-rapidity $v_1$-slope splitting ($\Delta \frac{dv_1}{dy}$) between $K^{*0}$ and $K^{+}$ in Au+Au collisons at $\sqrt{s_{NN}} = 27$ GeV in Fig.~\ref{fig4}. Notably, the magnitude of splitting is least in peripheral collisions and gradually increases towards the central collisons. In peripheral collisions, the duration of the hadronic phase life time is relatively shorter, alongside a reduced production of pions. Consequently, this leads to a lesser afterburner effect in the hadronic phase. Conversely, in central collisions, the hadronic phase persists for a more extended period, thus allowing a more pronounced hadronic afterburner influence.

We have further examined the centrality trend of the mid-rapidity $\frac{dv_1}{dy}$ splitting between $\phi$ meson and $K^{+}$ in Fig~\ref{fig4}. This investigation is particularly more illuminating as the $v_1$ of $\phi$ and $K^{+}$ appear to be less influenced by hadronic scatterings. Our findings reveal that the $\left[ \frac{dv_1}{dy} (\phi) -\frac{dv_1}{dy} (K^{+}) \right]$ consistently maintains a negative value across all centralities. This suggests that $\vert \frac{dv_1}{dy} (\phi)\vert > \vert \frac{dv_1}{dy} (K^{+})\vert $ holds true for all centralities which is an anticipated outcome stemming from the mass hierarchy. For comparison, we also depict the  $\left[ \frac{dv_1}{dy} (K^{*0}) -\frac{dv_1}{dy} (K^{+}) \right]$ for primordial hadrons, showing a consistently negative sign following the mass hierarchy. Interestingly, the sign of $\left[ \frac{dv_1}{dy} (K^{*0}) -\frac{dv_1}{dy} (K^{+}) \right]$ shifts to positive after the hadronic transport stage at all centralities. This is a clear signature of hadronic stage effect on $K^{*0}$ resonance which could be measured in experiments.

\begin{figure}
 \begin{center}
  \includegraphics[scale=0.5]{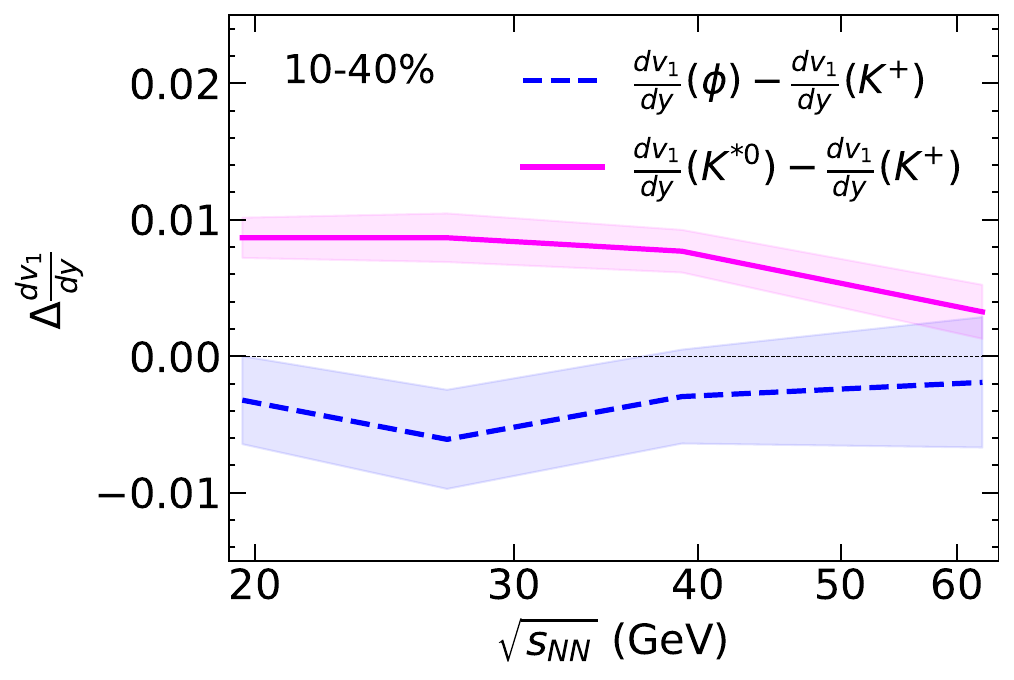}
 \caption{ The splitting of directed flow slope ($\frac{dv_1}{dy}$) between $K^{*0}$ and $K^{+}$, along with $\phi$ and $K^+$ has been plotted as a function of collision energy($\sqrt{s_{NN}}$) for Au+Au collisions of 10-40$\%$ centrality. }
 \label{fig5}
 \end{center}
\end{figure}

The collision energy dependence of the splitting of directed flow slope between $K^{*0}$ and $K^{+}$ has been plotted in Fig.~\ref{fig5} for 
Au+Au collisions of 10-40$\%$ centrality. It has been observed that the splitting is less in higher $\sqrt{s_{NN}}$ and it becomes more pronounced 
at lower collision energies owing to the larger tilt of the fireball at lower energies resulting in a larger asymmetric hadronic fireball at lower energies affecting $v_1$ of $K^{*0}$ more strongly. For comparison, we have also plotted the splitting between $\phi$ and $K^+$ that remains constant $\sim-0.005$.
Note that at even lower energies baryon stopping physics giving rise to non-trivial dynamics of conserved charges that could affect the directed flow of these resonances, particularly  that of $K^{*0}$ as it carries strangeness.

\section{SUMMARY}
\label{summ_sec}

In this work we have studied the effect of late stage hadronic interaction on the rapidity dependent directed flow ($v_1$) of $K^{*0}$ in a hybrid (hydrodynamics+hadronic transport) framework.
The study compares the $v_1$ results calculated from the hadrons which are directly produced from the hypersurface and resonance decay with those that undergo hadronic transport. The analysis reveals that the $v_1$ of $K^{*0}$ is strongly affected during the hadronic stage due to asymmetric signal loss in different sides of the $p_x$-axis in momentum space caused by the tilted fireball. We have also analysed the flow of $\phi$ for reference. We find that owing to small crosssection and long lifetime, flow of $\phi$ is unaffected by the hadronic afterburner.

The centrality dependence of the effect of hadronic interaction on $v_1$ of $K^{*0}$ has been studied by plotting the splitting of $v_1$ bewteen $K^{+}$ and $K^{*0}$
as a function of centrality at $\sqrt{s_{NN}}=27$ GeV. 
Notably, the observed splitting is minimal in peripheral collisions and progressively increases as collisions become more central. This trend signifies that the impact of hadronic interactions is most significant in central collisions, which can be attributed to the relatively longer duration of the hadronic phase and the comparatively higher yield of produced hadrons.
Furthermore, the beam energy dependence of the $v_1$ splitting between $K^{+}$ and $K^{*0}$ has been studied within the 10-40$\%$ centrality range. Interestingly, the effect of hadronic interactions becomes more pronounced in collisions with lower energy. This can be attributed to the larger tilt in the initial fireball resulting in an increasingly  asymmetric distribution of the hadronic medium in coordinate and momentum space. Thus, a comprehensive and quantitative investigation of the flow coefficients associated with $K^{*0}$ and other resonances, along with the model to experimental data comparison, holds the potential of yielding deeper insights into the nature of the hadronic phase.

\bibliographystyle{apsrev4-1}
\bibliography{hybrid}

\end{document}